\documentclass[aps,prb,twocolumn,groupedaddress,showpacs,floatfix]{revtex4}
\usepackage{graphicx}

\usepackage{hyperref}

\usepackage[ansinew]{inputenc}

\begin{document}

\title{Energy Level Lifetimes in the Single-Molecule Magnet $\rm Fe_8$: Experiments and Simulations}

\author{S. Bahr$^1$, K. Petukhov$^1$\footnote{presently at Physikalisches Institut III, Universit$\rm \ddot{a}$t Erlangen-N$\rm \ddot{u}$rnberg, Erwin-Rommel-Strasse 1, D-91058 Erlangen, Germany}, V. Mosser$^2$, W. Wernsdorfer$^1$}

\affiliation{
$^1$Institut N\'eel, CNRS/UJF, BP 166, 38042 Grenoble Cedex 9, France\\
$^2$Itron France, 76 avenue Pierre Brossolette, 92240 Malakoff , France}

\date{\today}

\begin{abstract}
We present pump-probe measurements on the single-molecule magnet $\rm Fe_8$ with microwave pulses having a length of several nanoseconds.
The microwave radiation in the experiments is located in the frequency range between 104~GHz and 118~GHz.
The dynamics of the magnetization of the single $\rm Fe_8$ crystal is measured using micrometer-sized Hall sensors.
This technique allows us to determine the level lifetimes of excited spin states, that are found to be in good agreement with theoretical calculations.
The theory, to which we compare our experimental results, is based on a general spin-phonon coupling formalism, which involves spin transitions between nearest and next-nearest energy levels. 
We show that good agreement between theory and experiments is only obtained when using both the \mbox{$\Delta m_S=\pm 1$} transition as well as \mbox{$\Delta m_S=\pm 2$}, where $\Delta m_S$ designates a change in the spin quantum number $m_S$.
Temperature dependent studies of the level lifetimes of several spin states allow us finally to determine experimentally the spin-phonon coupling constants.

\end{abstract}


\maketitle

\section{Introduction}

Single-molecule magnets (SMMs) are a novel class of materials, where almost identical magnetic molecules are regularly assembled in large crystals. 
Every molecule contains a well-defined structure of superexchange-coupled magnetic metal ions.
At low temperature the relative orientation of the strongly coupled, intramolecular spins are locked and each molecule can be described by one single, collective spin S. 
This approximative description of the molecule is called \emph{giant spin approximation} \cite{sessoli:jacs1993,sessoli:1993,novak:1995,barra:1996,aubin:jacs1998,caciuffo:1998}.
SMMs have attracted much research interest in the last decade because of their quantum properties that can be studied even on a macrosopic scale \cite{friedman:1996,thomas:1996,Sangregorio:1997,wernsdorfer:science1999,sorace:2003}.
In particular, quantum mechanical effects, such as quantum tunnelling of magnetization through the magnetic anisotropy barrier revealed the potential use of SMMs in the field of quantum computation and data storage \cite{leuenberger:2001,troiani:2005,ardavan:2007}.
SMMs are also a convenient model system to study the dynamics and the interaction of a well defined spin system with its environment.
Especially, interactions with nuclear spins or lattice vibrations in the crystal are a source of decoherence and therefore of particular interest \cite{stamp:1998prl,wernsdorfer:1999prl,stamp:prb2004,morello:2006}. 
The use of microwave radiation that induces transitions between different spin states can provide a direct access to the spin dynamics of the molecules.
Such experiments employing continuous-wave electron spin resonance \cite{barra:1996,hill:1998prl,mukhin:2001,zipse:2003} and pulsed microwaves time-resolved magnetometry \cite{sorace:2003,wernsdorfer:epl2004,bal:2004,barco:2004,petukhov:2005,bal:epl2005,cage:MMM2005,bal:JAP2006,bal:2007,petukhov:2007,loubens:2007jap,bahr:2007prl} were performed during last years in order to get a better understanding of the spin relaxation time $T_1$ and decoherence time $T_2$. 
These parameters were determined mainly by the use of detailed linewidth analysis of absorption peaks or direct relaxation measurements in the time domain \cite{sorace:2003,wernsdorfer:epl2004,bal:2004,barco:2004,petukhov:2005,bal:epl2005,cage:MMM2005,bal:JAP2006,bal:2007,petukhov:2007,loubens:2007jap,bahr:2007prl}.
A possibility to overcome various experimental difficulties (like extensive heating of the sample by long microwave pulses, phonon-bottleneck effects, and thermal relaxation) can be the use of very short microwave pulses or sequences of pulses on a nanosecond scale.

Concerning fast spin relaxation times of the order of several hundreds of nanoseconds we recently proposed a pump-probe technique employing a sequence of microwave pulses to resolve the basic mechanisms of spin-phonon coupling in $\rm Fe_8$ \cite{bahr:2007prl}.
This kind of technique employs very short microwave pulses and therefore thermal heating of the material is very small.
The time resolution of the dynamical processes obtained pump-probe experiments can be as fast as several nanoseconds and as a consequence we think they are much more precise and appropriate than direct relaxation measurements in the time domain.


In this paper we study the spin dynamics of the SMM $Fe_{8}O_{2}(OH)_{12}(tacn)_{6}$, hereafter called Fe$_8$ \cite{wieghardt:1984}.
This molecule contains eight Fe(III) ions with spins $s=5/2$. These
spins are strongly superexchange coupled forming a spin ground
state $S=10$ and the molecule can be described at low temperature by an effective Hamiltonian \cite{barra:1996}
\begin{equation}
\mathcal{H} = - DS_z^2 + E(S_x^2 -S_y^2) + \mathcal{O}(4) +
g\mu_{\rm B}\vec{S}\cdot\vec{H}
\end{equation}

\begin{figure}[t]
\includegraphics[width=3.2in]{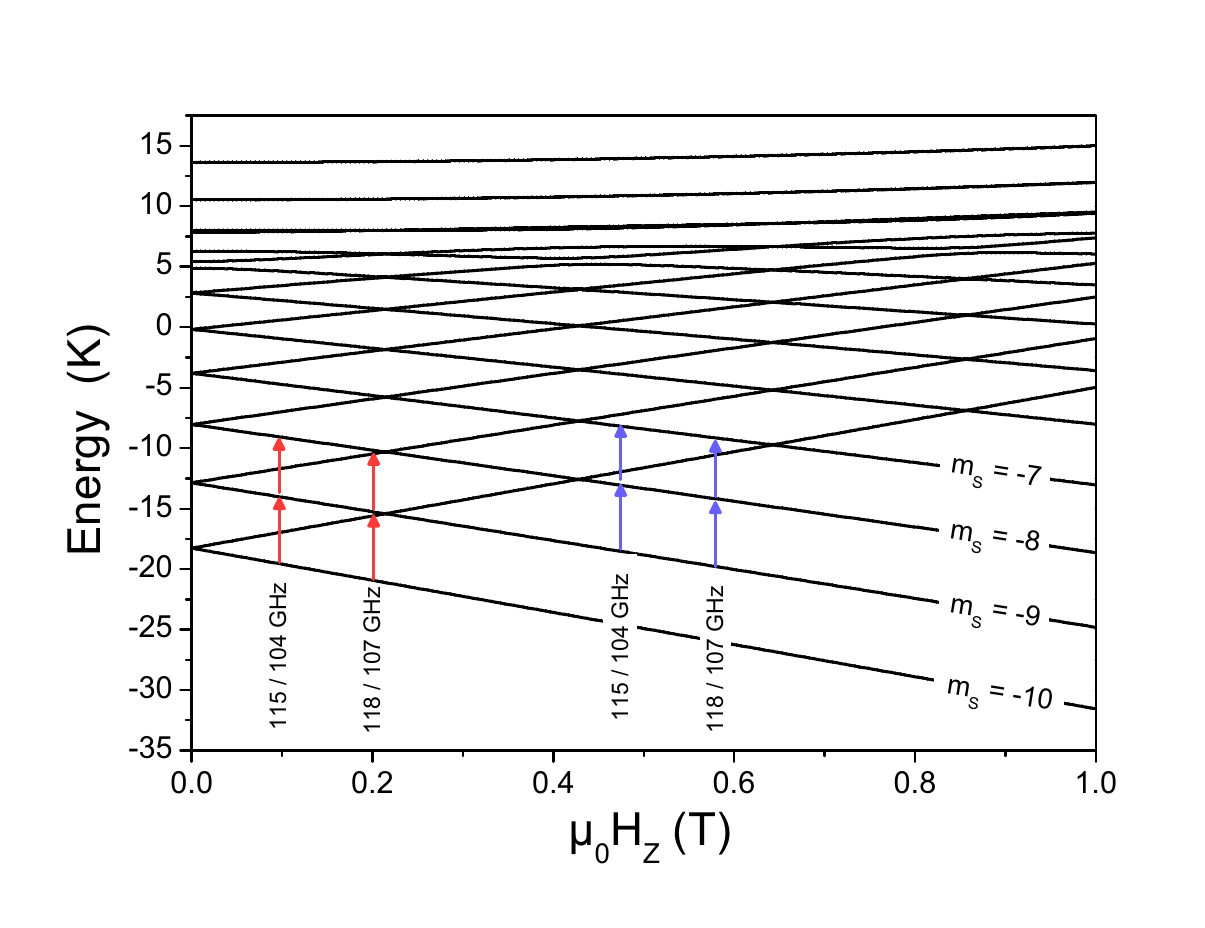}
\caption{\label{fig1}(Color online) Zeeman diagramm of the SMM Fe$_8$ with some microwave induced transitions between spin states. For a certain given combination of two microwave frequencies, i.e. 104~GHz and 115~GHz, one can find a magnetic field value where the higher microwave frequency excites spins from a level $m_S$ to $m_S+1$ and the lower frequency is in resonance between $m_S+1$ and $m_S+2$.}
\end{figure}
$\vec{H}$ is the applied magnetic field, $\mathcal{O}(4)$ contains
fourth order terms of spin operators, $g\approx 2$ represents
the gyromagnetic factor and $\mu_{\rm B}$ denotes the Bohr magneton.
The anisotropy parameters $D=0.275$~K and $E=0.046$~K have been determined by various experimental methods, such as high-frequency EPR techniques, frequency domain magnetic resonance spectroscopy and neutron spectroscopy and they yield  similar results.
\cite{barra:1996,wernsdorfer:science1999,mukhin:2001,caciuffo:1998,park:2002} 
The non-diagonal terms in the Hamiltonian are responsible for the
tunneling processes between spin states, whereas $D$ defines the
anisotropy barrier of approximately 25~K.

This paper presents pump-probe measurements on the SMM Fe$_8$.
We focus our work on the level lifetimes of different spin states as a function of temperature and applied magnetic field.
In Section II we describe the experimental setup and the principle of the pump-probe technique. 
In Section III we present the experimental results of the pump-probe experiments that will be compared in Section IV to simulations based on a general spin-phonon coupling theory.
The experimental data are discussed in Section V and Section VI gives finally some concluding remarks.

\section{Experimental techniques}

The measurements are performed using a commercial 16~T
superconducting solenoid and a cryostat in the temperature 
range of 1.4~K to 15~K with high temperature stability.
The magnetization of the Fe$_8$ sample is measured with a
Hall magnetometer. 

The Hall bars were patterned by Thales Research and Technology (Palaiseau),
using photolithography and dry etching, in a delta-doped AlGaAs/InGaAs/GaAs
pseudomorphic heterostructure grown by Picogiga International using
molecular beam epitaxy (MBE). A two-dimensional electron gas is induced in
the 13 nm thick $In_{0.15}Ga_{0.85}As$ well by the inclusion of a Si delta-doping
layer in the graded $Al_xGa_{1-x}As$ barrier. 
All layers, apart from the quantum well, are fully depleted of electrons and holes. The two-dimensional electron gas density $n_s$ is about $8.9 \times 10^{11}~cm^{-2}$ in the quantum well, corresponding to a sensitivity of about 700~$\Omega$/T, essentially constant below $-100~^\circ$C. 

The sample is placed on top of the 10~$\mu$m $\times$ 10~$\mu$m Hall junction
with its easy axis approximately parallel to the magnetic field of
the solenoid.
The Hall voltage being about proportional to the magnetization and of the order of magnitude of several $\rm \mu$V, is amplified by a low noise preamplifier with large frequency bandwidth $f_{BW}>100$~MHz.
The time-resolved magnetization measurements were averaged typically over 1000 events.

\begin{figure}[t]
\includegraphics[width=3.2in]{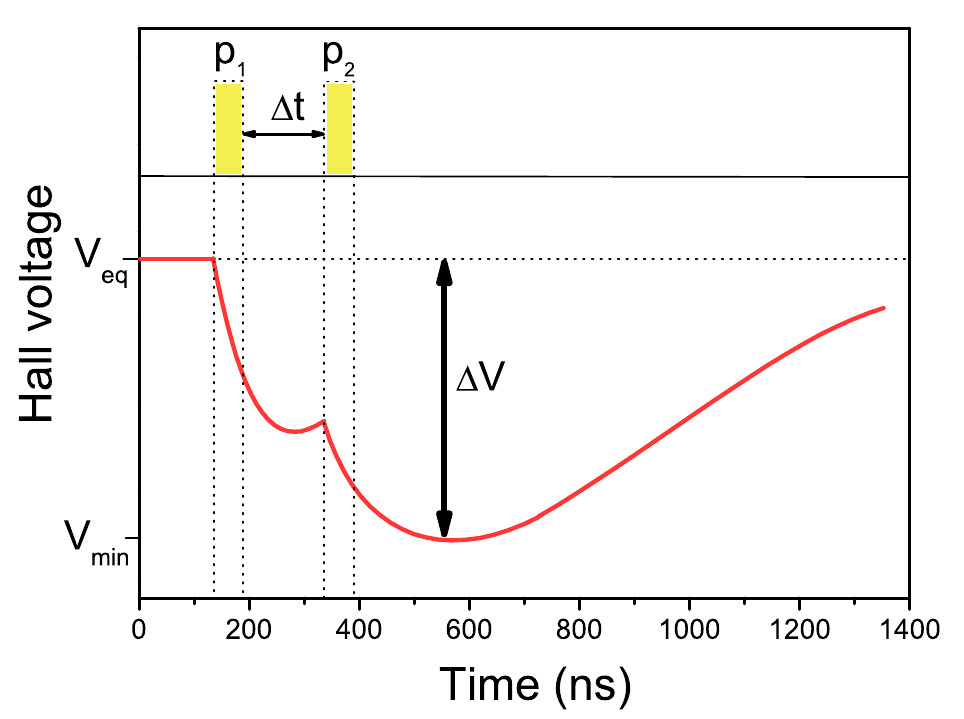}
\caption{\label{fig2}(Color online) Sketch of a typical pump-probe experiment at low temperature. 
Two microwave pulses $\rm p_1$ and $\rm p_2$ with different frequencies and pulse lengths of typically less than 50~ns are separated by a delay $\Delta t$.
They excite the spin system and the magnetization of the sample, i.e. the Hall voltage decreases, reaches a minimum and relaxes back to the equilibrium value.
The Hall voltage amplitude $\Delta V = V_{eq}-V_{min}$ depends on the amount of spins excited by the two microwave pulses.}
\end{figure}

Microwaves are generated by two continuous wave,
mechanically tunable Gunn oscillators in a frequency range of 97~GHz to 119~GHz. 
Pulses are generated using two SPST fast PIN diode switches (switching time less of
than 3~ns) triggered by a commercial pulse generator. 
An oversized circular waveguide of 10~mm diameter leads the microwaves into the
cryostat to the sample.

\begin{figure}[b]
\includegraphics[width=3.2in]{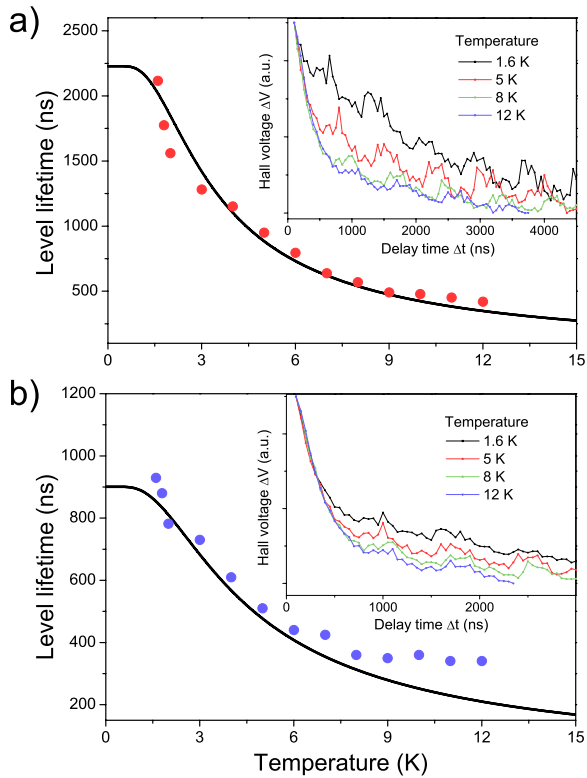}
\caption{\label{fig3}(Color online) Experimental results of the pump-probe experiments with frequencies 104~GHz and 115~GHz. The pump pulse has a length of 40~ns, whereas the probe pulse has a length of 20~ns.
(a) Measurements at a magnetic field of $\mu_0 H_z = 0.12$~T, i.e. spin transitions between the states $m_S=-10$, $m_S=-9$ and $m_S=-8$.
The inset shows $\Delta V$ as a function of the delay time between the two pulses for different temperatures in the range of 1.6~K to 12~K.
The characteristic relaxation time of $\Delta V(\Delta t)$ is shown as a function of the temperature.
(b) Measurements at a magnetic field of $\mu_0 H_z = 0.48$~T, i.e. spin transitions between the states $m_S=-9$, $m_S=-8$ and $m_S=-7$.
The inset shows $\Delta V$ as a function of the delay time between the two pulses for different temperatures in the range of 1.6~K to 12~K.
The characteristic relaxation time of $\Delta V(\Delta t)$ is shown as a function of the temperature.
We find good agreement with theoretical calculation (black lines).
}
\end{figure}

The pump-probe technique uses a sequence of two microwave pulses with pulse lengths as short as 20~ns.
The two microwave pulses $p_1$ and $p_2$ are delayed in time and they excite spins from different levels $m_S$ (Fig.~\ref{fig1}).
In general, the pump-probe experiment involves three energy levels $m_S$, $m_S+1$ and $m_S+2$.
The resonant frequency between $m_S$ and $m_S+1$ as well as $m_S+1$ and $m_S+2$ depend on the applied magnetic field $\mu_0 H_z$.
We carried out pump-probe experiments at four different magnetic field values using two different pairs of microwave frequencies.
At a magnetic field $\mu_0 H_z = 0.12$~T ($\mu_0 H_z = 0.48$~T) we used the resonant frequencies $f=104$~GHz and $f=115$~GHz to study the level lifetime of the level $m_S=-9$ ($m_S=-8$).
At a magnetic field $\mu_0 H_z = 0.2$~T ($\mu_0 H_z = 0.56$~T) we used the resonant frequencies $f=107$~GHz and $f=118$~GHz to study the level lifetime of the level $m_S=-9$ ($m_S=-8$).
During the pump-probe experiments the magnetization of the sample, which is directly proportional to the Hall voltage, is recorded as a function of time for fixed delay $\Delta t$ between the two microwave pulses (Fig.~\ref{fig2}).
The Hall voltage V is a precise measure of the magnetization of the sample.
As the temperature in the crystal is hardly affected by the very short microwave pulses, we can suppose that thermal activation of spins to higher levels is very weak.
Therefore $\Delta V$, as defined in figure 2, is proportional to the number of spins excited by the two microwave pulses to higher energy levels.
In particular, the decrease of the magnetization due to the second microwave pulse $p_2$ is proportional to the amount of spins in the level $m_S+1$.
When plotting the magnetization decrease, which is proportional to $\Delta V$, as a function of the delay time $\Delta t$ between the two microwave pulses, we get information about the depopulation of the level $m_S+1$ and therefore we can determine the level lifetime of the level $m_S+1$.
This decay of the population of the level $m_S+1$ is mainly due to spin-phonon interaction in the temperature range of our experiments.

\section{Measurements}

We already showed the possibility to investigate level lifetimes with the above described pump-probe technique elsewhere.\cite{bahr:2007prl}
This present publication is a substantial extension of previous work.
In particular we compare pump-probe measurements at several magnetic fields and with different microwave frequencies.

The insets of figure~\ref{fig3} show the Hall voltage decrease $\Delta V$ as function of the delay $\Delta t$ between the microwave pulses with frequencies $f_1=115$~GHz and $f_2=104$~GHz.
The relaxation of $\Delta V$ is fitted with a single exponential function and the characteristic relaxation time is plotted as a function of the temperature in the range 1.6~K to 12~K.
The relaxation time depends on the spin level involved in the experiment.

When working at a magnetic field $\mu_0 H_z = 0.12$~T we study the level lifetime of $m_S=-9$ (Fig. \ref{fig3}a)
The relaxation time is at low temperature around 2~$\rm \mu$s, whereas for higher temperatures it is around 500~ns.
For the level $m_S=-8$ at a magnetic field $\mu_0 H_z = 0.48$~T we observe at low temperature a relaxation time of 1~$\rm \mu$s and for higher temperatures less than 400~ns (Fig. \ref{fig3}b).

\begin{figure}[t]
\includegraphics[width=3.2in]{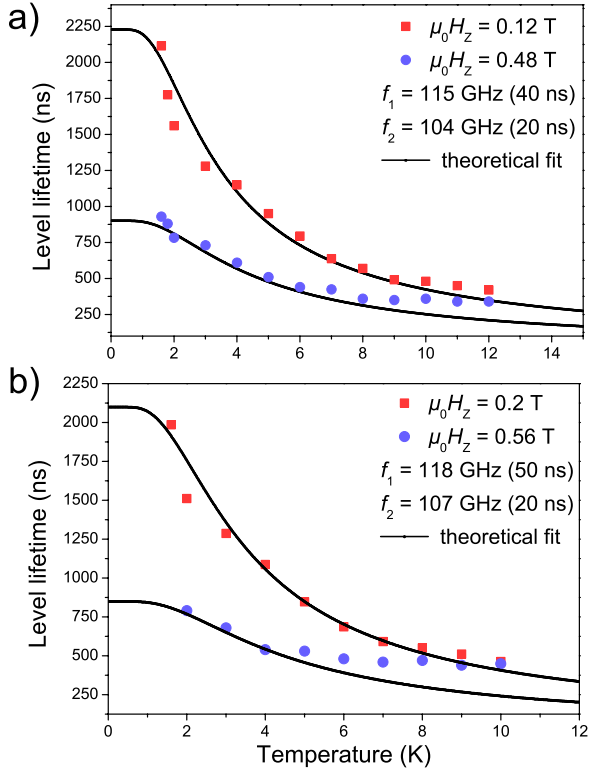}
\caption{\label{fig4}(Color online) Summary of various different pump-probe measurements employing two different pairs of microwave frequencies.
(a) With the combination 104~GHz/115~GHz the level lifetime of $m_S=-9$ can be probed at $\mu_0 H_z=0.12$~T, whereas the lifetime of the $m_S=-8$ states can be probed at $\mu_0 H_z=0.48$~T.
(b) The combination 107~GHz/118~GHz allows to determine the level lifetime of $m_S=-9$ and $m_S=-8$ at magnetic fields of $\mu_0 H_z=0.2$~T and $\mu_0 H_z=0.56$~T, respectively.
The level lifetime measurements are in good agreement with theoretical simulations (black lines).
}
\end{figure}

Figure~\ref{fig4} shows two generic plots of the relaxation times as a function of temperature and the four different magnetic field positions as shown in figure~\ref{fig1}.
In figure~\ref{fig4}a the relaxation time is plotted as a function of the temperature for the microwave frequencies $f_1=115$~GHz and $f_2=104$~GHz.
The measurements on the spin level $m_S=-9$ yield significantly longer relaxation times than measurements on the level $m_S=-8$.

In figure~\ref{fig4}b the relaxation time is plotted as a function of the temperature for the microwave frequencies $f_1=118$~GHz and $f_2=107$~GHz.
The results are similar to the ones obtained with $f_1=115$~GHz and $f_2=104$~GHz.
Note that in particular the lifetime of the spin level $m_S=-9$ is significantly longer than the lifetime of the level $m_S=-8$.

All experimentally obtained level lifetimes are compared to a theoretical model based on general spin-phonon coupling theory.
The theoretical calculations of the level lifetimes that are shown in the figures~\ref{fig3} and \ref{fig4} (black lines) are performed with one set of parameters, that will be detailed hereafter in the theoretical section.

\section{Theoretical approach}

We follow in this section the theoretical description of spin-phonon coupling in molecular high spin systems first described by \emph{Villain et al.} and later extended by \emph{Garanin et al.} and \emph{Leuenberger et al.}\cite{villain:1994,garanin:1997,loss:prb2000}

In general, the spin-phonon coupling can induce transitions of the form $\Delta m_S=\pm 1$ and $\Delta m_S=\pm 2$ of the projection of the spin along the z direction, where $\Delta m_S$ denotes the change of the spin quantum number $m_S$.
For a given spin state $m$ we obtain therefore four different rate equations concerning phonon induced transitions $W_{m \rightarrow m\pm 1}$ and $W_{m \rightarrow m \pm 2}$ to higher- and lower-lying energy levels.
A theoretical approach using a generalized master equation which treats phonon-induced spin transitions leads to explicit formulas of these spin-phonon processes \cite{loss:prb2000}
\begin{eqnarray}
\label{eq1}
W_{m \rightarrow m \pm 1} = \frac{g^2 s_{\pm 1}}{12 \pi \rho c^5 \hbar^4}\frac{(\epsilon_{m \pm 1}-\epsilon_m)^3}{e^{\beta(\epsilon_{m \pm 1}-\epsilon_m)}-1} \\ 
\label{eq2}
W_{m \rightarrow m \pm 2} = \frac{17 g^2 s_{\pm 2}}{192 \pi \rho c^5 \hbar^4}\frac{(\epsilon_{m \pm 2}-\epsilon_m)^3}{e^{\beta(\epsilon_{m \pm 2}-\epsilon_m)}-1} 
\end{eqnarray}
where $g$ denotes the spin-phonon coupling constant, $s_{\pm 1}=(s\mp m)(s\pm m+1)(2m \pm 1)^2$, 
$s_{\pm 2}=(s\mp m)(s\pm m+1)(s\mp m-1)(s\pm m+2)^2$, $\rho=1.9\cdot10^3 kg/ m^{3}$, $\beta=1/k_B T$ and $c$ is the speed of sound.
It has been shown, that the spin-phonon coupling constants $g_i$ are all of the same order of magnitude and their magnitude has shown to be similar to the anisotropy energy $D$ in comparable molecular systems.\cite{loss:prb2000}

\begin{figure}[t]
\includegraphics[width=3.0in]{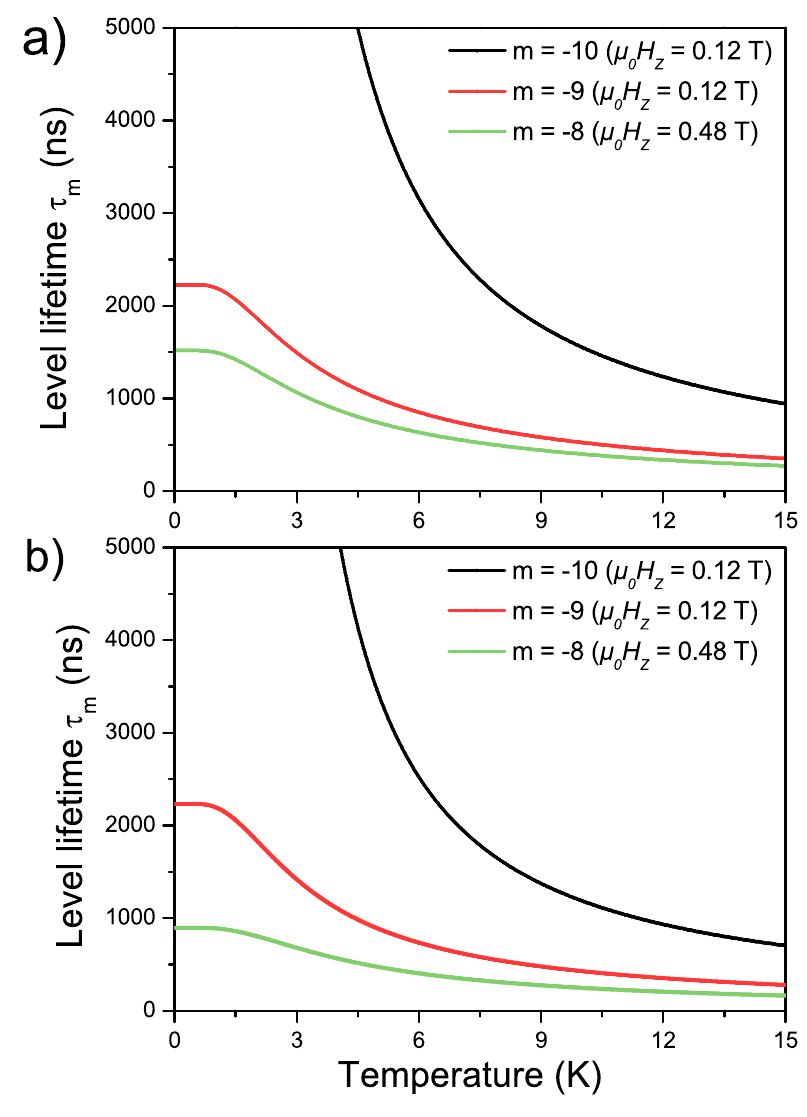}
\caption{\label{fig5}(Color online) Theoretical calculations of the level lifetimes of three different spin states $m=-10$, $m=-9$ and $m=-8$. The magnetic field $H$ is chosen according to our pump-probe experiments with 115~GHz and 104~GHz (Fig.~\ref{fig4}a).
In figure (a) we only considered the $\Delta m_S=\pm 1$ process, whereas in figure (b) we calculated the level lifetimes according to equation (\ref{tau}) using the $\Delta m_S=\pm 1$ and $\Delta m_S=\pm 2$ spin-phonon processes. 
Note, that the level lifetimes of $m=-9$ and $m=-8$ shown in figure 5b correspond to the theoretical simulation in figure 4a.
}
\end{figure}

Finally, the lifetime $\tau_m$ of a spin level $m$ is defined by 

\begin{eqnarray}
\label{tau}
1/\tau_m &=& W_{m \rightarrow m+1} + W_{m \rightarrow m-1} \nonumber\\
&+& W_{m \rightarrow m+2} + W_{m \rightarrow m-2}
\end{eqnarray}

The lifetimes of the different spin levels $m$ show large differences at low temperature.
The magnitude of the level lifetimes of excited levels is much smaller than the ground state level lifetime.
Additionally, all the level lifetimes exhibit a strong temperature dependence.

In figure \ref{fig5} several level lifetimes are plotted as a function of the temperature. 
In the two subfigures we compare directly the influence of the $\Delta m_S = \pm 2$ process on the different level lifetimes.
As can be seen in equations (\ref{eq1}) and (\ref{eq2}), the level lifetimes are very sensitive to variation of the speed of sound $c$ or the spin-phonon coupling constants.
They also depend strongly on the relative magnitude of the $\Delta m_S = \pm 1$ and $\Delta m_S = \pm 2$ processes.
In fact, the $\Delta m_S = \pm 2$ process gives a big contribution to the excited level lifetimes and at low temperature this process strongly influences the spin dynamics and shortens the level lifetimes.
This property can be seen clearly when comparing the level lifetimes of the excited spin state $m=-8$ in figures \ref{fig5}a and \ref{fig5}b.
The additional $\Delta m_S = \pm 2$ process reduces the level lifetimes at low temperature by nearly a factor 1.5.

\section{Discussion}

The presented pump-probe experiments using nanosecond microwave excitations allow us to determine directly energy level lifetimes.
Figures 3 and 4 show the obtained level lifetimes of different energy levels at different magnetic fields and as a function of the temperature.
The obtained energy level lifetimes for the levels $m=-9$ and $m=-8$ are in the range between 2.2~$\mu$s and 300~ns and
the dependence on temperature and magnetic field can be well reproduced by general spin-phonon coupling theory.
However, for higher temperatures ($T>8$~K) we observe deviations from the theoretical calculations, especially for the measurements on the $m=-8$ state.
This might be due to the giant spin approximation, i.e. due to disregarding of higher spin multiplets, that start to get populated at higher temperatures (the $S=9$ multiplet is located at about $\Delta E \approx 24$~K above the ground state multiplet \cite{zipse:2003}).

In addition, the level lifetimes seem not to be influenced very much by the presence of a proximate level anticrossing (as in the case of measurements at 0.2~T)~\cite{bahr:2007prl}.
In fact, one could have supposed, that the relaxation of magnetization gets accelerated when additional relaxation paths get available.
However, we do not notice any dramatically changes in the level lifetimes in our experiments, when measuring near a level anticrossing.
In consequence, the dominant relaxation mechanism seems to be based on spin-phonon interaction in the temperature range we investigated.

In particular, the prefactor of the simplified spin-phonon rate equations (\ref{eq1}) and (\ref{eq2}) depends on the spin-phonon coupling constant and on the transverse speed of sound in the bulk material.
Unfortunately, our measurements do not allow to fit both of these parameters independently.
The speed of sound in $\rm Fe_8$ has been studied by several techniques, such as specific heat measurements and NMR.\cite{gomes:1999condmat,furukawa:prb2001fe8,mettes:2001prb,evangelisti:2005prl}
The published results differ a lot and range from $c=670 \frac{m}{s}$ to $c=1500 \frac{m}{s}$.
Nevertheless, our measurements allow to determine the ratio $\frac{g^2}{c^5}$, where $g$ is the spin-phonon coupling constant and $c$ the speed of sound. 
We find a value of $\frac{g^2}{c^5}\approx 1\cdot 10^{-17}\frac{K^2 s^5}{m^5}$, which leads to a spin-phonon coupling constant $g$ in the range between 0.036~K and 0.275~K (when $c$ is chosen in the range mentioned above).

Recent publication in the field suggest to describe the spin-phonon interaction only on the basis of the 
\mbox{$\Delta m_S=\pm 1$} process and to neglect completely the \mbox{$\Delta m_S = \pm 2$} processes.\cite{bal:2007}
This will result in substantially longer level lifetimes of the excited spin states and therefore slow down the spin dynamics.
In particular, figure \ref{fig5} shows that the $\Delta m_S=\pm 2$ spin-phonon process has great influence in the low temperature behavior of the level lifetimes. 
Especially for the level $m=-8$ and higher excited levels this supplementary process gives rise to a significant reduction of the level lifetime compared to the $m=-9$ level. When only taking $\Delta m_S=\pm 1$ spin phonon relaxation process, the levels $m=-9$ and $m=-8$ the difference of their level lifetimes at low temperature decreases significantly.
Our experimental results are only well reproduced by spin-phonon coupling theory when using the $\Delta m_S = \pm 1$ and $\Delta m_S = \pm 2$ processes with the choice of parameters as explained above.

\section{Conclusion}

In conclusion we presented pump-probe experiments on the SMM $\rm Fe_8$ with pulsed microwaves on a nanosecond scale at frequencies in the range of 104~GHz to 118~GHz.
These experiments allowed us to investigate the level lifetimes of excited spin states as a function of temperature and applied magnetic field.
Typically, we observe level lifetimes of the order of magnitude of 400~ns up to more than 2~$\mu$s.
The temperature depended measurements presented for the $m=-9$ and $m=-8$ energy levels can be explained by a general spin-phonon coupling theory.
The fit of our experimental data are consistent with published values for the speed of sound and spin-phonon coupling constants.

\begin{acknowledgments}
We acknowledge motivating discussions with J. R. Friedman and E. M. Chudnovsky.
The sample for the investigations was kindly provided by A.~Cornia. 
This work is partially financed by EC-RTN-QUEMOLNA Contract No.
MRTN-CT-2003-504880 and MAGMANet. 
The authors thank E. Eyraud, J.-S. Pelle and J. Florentin for their useful technical contributions to this work and A.-L. Barra for scientific support.
\end{acknowledgments}

\end{document}